\documentclass[%
 reprint,
 amsmath,amssymb,
 aps, superscriptaddress
]{revtex4-2}

\usepackage{graphicx}
\usepackage{dcolumn}
\usepackage{bm}
\usepackage{hyperref}
\hypersetup{colorlinks=true, citecolor=blue, urlcolor=blue, linkcolor=blue}

\begin{document}

\preprint{APS/123-QED}

\title{Laser Systems for High Fidelity Control and Entanglement of Neutral Atomic Qubits}
\author{C.J. Picken}
\email{craig.picken@m2lasers.com}
\author{I. Despard}%
\author{A. Kelly}%
\affiliation{%
 M Squared Lasers Limited\\ 1 Kelvin Campus, West of Scotland Science Park,\\
Glasgow, G20 0SP, United Kingdom
}%

\author{J.D. Pritchard}
\affiliation{
 SUPA, Department of Physics, University of Strathclyde, \\Glasgow G4 0NG, United Kingdom
}

\author{J.R.P. Bain}%
\author{N. Hempler}%
\author{G.T. Maker}%
\author{G.P.A Malcolm}%

\affiliation{%
 M Squared Lasers Limited\\ 1 Kelvin Campus, West of Scotland Science Park,\\
Glasgow, G20 0SP, United Kingdom
}%

\date{\today}

\begin{abstract}
We present new photonics and electronics packages recently developed by M Squared Lasers specifically tailored for scalable neutral atom quantum computing; a high power 1064~nm system for scalable qubit number, a phase locked system for high fidelity single qubit control, and robust cavity locked systems for high fidelity Rydberg operations. 
We attain driven coherence times competitive with current state-of-the-art for both ground state Raman and ground-Rydberg transitions without cavity filtering, providing an excellent platform for neutral atom quantum computing. These systems are benchmarked by creating entangled Bell states across 7 atom pairs, where we measure a peak raw fidelity of $F\ge0.88(2)$ and a peak SPAM corrected of $F_C\ge0.93(3)$ via a two-qubit $CZ$ gate.

\end{abstract}

\maketitle

\section{\label{sec:level1}Introduction}

The viability of quantum computing is underpinned by the coherent control of quantum systems. Several qubit technologies have demonstrated high fidelities approaching levels where quantum error correction could lead to real world applications in the near future~\cite{Egan:2021aa, Acharya:2022aa, Evered:2023aa}. Of these candidates, ion and neutral atom platforms rely on coherent light-matter interactions to manipulate quantum states and create entanglement~\cite{Levine:2018aa,Ballance:2016aa}. 

Recent progress has seen neutral atoms emerge as a promising candidate to realise a full scale quantum computing system, with demonstrations of single qubit gates which exceed the threshold for fault tolerant quantum computation over large arrays~\cite{Nikolov23} and significant increases in two qubit gate fidelities via Rydberg interactions~\cite{Levine:2019aa, Graham:2019aa, Evered:2023aa}. Neutral atom processors also provide advantageous properties to scale qubit numbers to the levels required for error correction~\cite{Wang:2020aa,Cong:2021aa,Graham:2019aa,Ebadi:2021aa,Schymik22,Singh22,Brown19,Sheng22,Tian:2022aa}, with systems up to 1000 qubit sites currently being demonstrated~\cite{Huft:2022aa} and demonstration of small scale digital algorithms~\cite{Graham:2022aa, Bluvstein:2022aa}. The flexibility of neutral atoms in reconfigurable atomic tweezers also allows such platforms to be used for analogue quantum computing and quantum simulation~\cite{Henriet:2020aa}, with recent demonstrations of superlinear quantum speedup~\cite{Ebadi:2022aa} which can be extended to a wide range of combinatorial optimisation problems~\cite{Nguyen:2022aa}.

\begin{figure}[t!]
\includegraphics{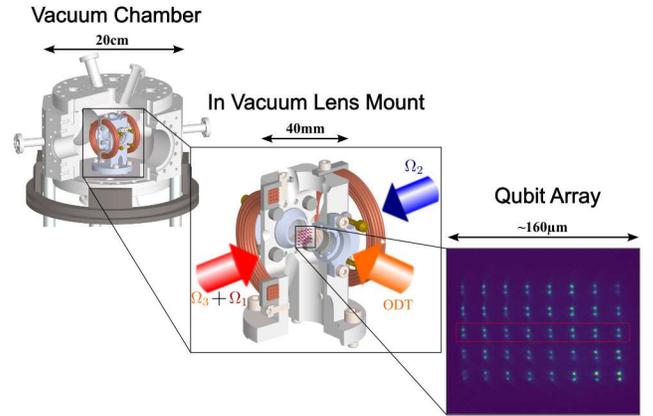}
\label{fig:fig1}
\caption{Outline of our setup - a reconfigurable qubit array is created at the focus of in-vacuo lenses by a 1064 nm optical dipole trap (ODT). Global beams perpendicular to the dipole trap and along the quantisation axis enable qubit operations. Shown is an average image of 40 pairs of atoms. The work presented in this paper focuses on the central row highlighted by the red box to suppress variations coming from spatial uniformity of the Rydberg excitation lasers.}
\end{figure}

Detailed analysis highlights the critical need to overcome limiting noise factors from laser sources to realise high fidelities~\cite{Leseleuc:2018aa, Jiang:2022aa}, with the low frequency noise of Ti:Sapphire based lasers offering a significant advantage over semiconductor diode lasers~\cite{Jiang:2022aa}. In this paper, we introduce a comprehensive photonics backbone of lasers used for qubit preparation, high fidelity control and entanglement, as part of our neutral atom quantum processor, Maxwell. All systems presented are controlled with our Ice Bloc interface.  Locking, including to our commercially available cavity, vapour cell, and to other SolsTis lasers are handled by proprietary M Squared electronics. We first review the experimental setup and laser systems, and then provide experimental demonstration of extended coherence time in Rydberg excitation and high-fidelity entanglement generation.

\section{\label{sec:level1}Experimental Setup}

Our setup uses seven M Squared laser systems, for simplicity we will split the systems into groups and discuss in further detail. The geometry of our qubit beams are shown in Fig~1 and each laser system depicted in Fig~2(a).

\begin{figure}[t!]
\includegraphics{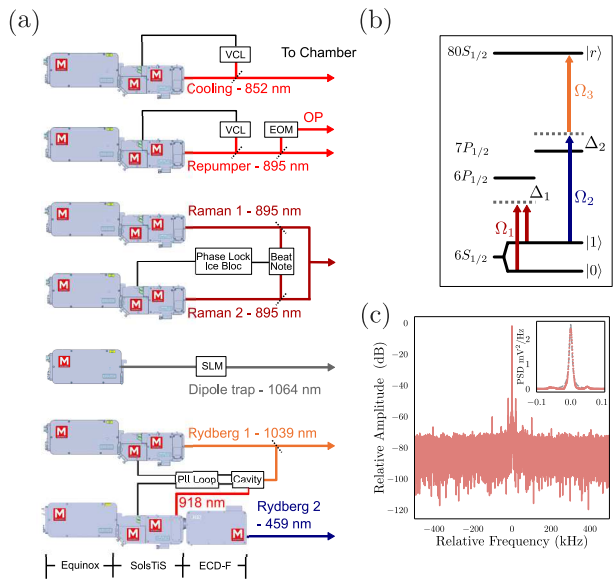}
\label{fig:fig2}
\caption{(a) Laser systems used in this work: 852 and 895~nm SolsTiS PI used for cooling, repumping and optical pumping. Two phase locked 895~nm SolsTiS PI systems used for Raman transitions. A 1064~nm Equinox for atom trapping. Cavity locked 1039 and 918~nm SolsTiS PI systems with an ECD-F module frequency doubling the 918~nm light for Rydberg excitation. (b) Level diagram for the qubit operations presented in this paper. (c) Beatnote between the Rydberg lasers both locked to independent axes of the cavity at 918 nm.}
\end{figure}

\subsection{Qubit Register}

A schematic of our apparatus for generating reconfigurable qubit arrays is shown in Fig.~1(a), which is based on the experimental design detailed in Ref.~\cite{Nikolov23}. Approximately 10$^5$ caesium atoms are trapped in a 3D magneto-optical trap (MOT) in 100 ms from a 2D MOT. Cooling and repump light are provided by two SolsTiS PI lasers at 852 nm and 895 nm respectively. Each laser is dither locked to saturation spectroscopy using M Squared's Vapour Cell Lock module.

Following loading of the MOT, the atoms are further cooled by polarisation gradient cooling (PGC). The atoms are then loaded into a 1064~nm holographic array created via a spatial light modulator (SLM)~\cite{Nogrette:2014aa}, with a $1/e^2$ waist of 1.55~$\mu$m achieved via in-vacuum high numerical aperture lens~\cite{Sortais22}. After loading into the traps, cooling light is applied to drive light assisted collisions for a duration of 50 ms, resulting in the probabilistic loading (55\%) of a single atom at each site~\cite{Schlosser02}. The diode-pumped, solid state 1064~nm system has been specifically engineered for quantum computing applications, with 20 W of single longitudinal and transverse mode power available. This provides enough power to facilitate the trapping of over 225 qubits with a 3~mK trap depth~\cite{Nikolov23}, accounting for losses of up to 50~\% between the laser output and the focus at the atoms. 

We verify site occupancy using the following imaging sequence. A single retro-reflected beam at a detuning of -2.5$\Gamma$ is applied for a duration of 20~ms. The imaging beam is chopped out of phase with the dipole traps (3~mK) at 1~MHz to remove associated light shifts from the trapping potentials. The fluorescence is collected by the in-vacuo lens and is imaged onto a sCMOS camera (Photometrics Prime BSI). Atom detection is performed by thresholding the measured count data, based on fitting the distribution from 200 loads~\cite{Picken:2017aa}. Imaging is then followed by another PGC phase before the trap depth is adiabatically ramped down to 250~$\mu$K, resulting in single atom temperatures of 5~$\mu$K measured using release-recapture~\cite{Tuchendler08}. The atoms are then prepared in the $|1\rangle = |F=4,m_{f}=0\rangle$ state using linearly ($\pi$) polarised 895 nm light from our repump laser with a sideband generated by an EOM, in a bias field of 7~G. This heats the atoms to a temperature of 10~$\mu$K.

At this stage, we perform our quantum manipulations of the atoms, and dependent on the experiment, a blowaway pulse is applied. The trap potentials are then ramped back to the imaging settings (3~mK) such that a second image can be taken to measure the outcome of the test performed. The lifetime in our vacuum chamber is measured to be 22.3(6)~s. 

For the measurements presented in this paper we use 8 pairs of atoms separated by $\approx 20~\mu$m and an interatomic separation of $\approx 6~\mu$m, with a slight variation in the separation occurring at trap sites diffracted furthest away from the zeroth order by the SLM. The atom retention between images 98.8-99.6~\% at each site with a mean across the array found to be  99.3(3)~\%, and represents our error in measuring if an atom is in F=3 in our state-selective readout (false-negative). The blowaway error of measuring an atom in F=4 (false-positive) at each site is 0.1-0.7~\% with a mean of 0.4(2)~\%. The variation in these benchmarks across our trapping sites is due to a inhomogeneity in trap depths across the array.

\subsection{Ground State Qubit Control}

Global single qubit operations are applied to the array using a 895~nm phase locked system detuned from the $6P_{1/2}$ $D_1$ line, which facilitate Raman transitions between $|1\rangle$ and $|0\rangle = |F=3, m_f =0\rangle$. Two SolsTiS PI systems each provide $>$~3.5 W of power. The primary and secondary lasers are phase locked at 8.92~GHz by feeding the beatnote of the two lasers into our Phase Lock Ice Bloc. This lock is maintained through three PI feedback loops to the secondary laser. A fast PI feeds back to the intracavity electro-optical modulator (EOM), slow feedback to the fast piezo (DC - 250 Hz) and a final feedback to the slow resonator peizo.  The slow resonator is controlled by a digital to analog convertor which is updated by the Ice Bloc digital signal processor software to maintain the fast piezo and EOM feedback voltages within their operating range, providing long-term locking stability. The frequency separation between the lasers is then controlled by a series of acoustic-optical modulators before the fibre to realise a separation of $\omega_{\rm{HFS}}= 9.193$~GHz.

\begin{figure}[t!]
\includegraphics{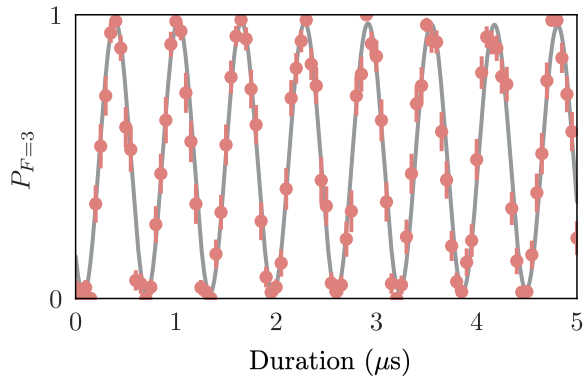}
\label{fig:fig3}
\caption{Raman Rabi oscillation of a single site using our phase locked Raman system, $\Omega_{1}/2\pi =$ 1.58(1) MHz at $\Delta_{1}/2\pi =$ -250 GHz. The standard deviation in Rabi frequency across our 16 sites is 0.5~\%}.
\end{figure}

The large power overhead provided by the SolsTiS allows for operation at large detunings from the intermediate $6P_{1/2}$ state, to suppress errors arising from excited state scattering. The available power also enables us to operate with large beams to suppress variation in Rabi frequency across the array. For the experiments presented, we typically use 400 mW total Raman power at the atoms and focus to a $1/e^2$ waist of 180~$\mu$m. With these parameters we obtain two-photon couplings of $\Omega_{1}/2\pi =$ 1.64 MHz at $\Delta_{1}/2\pi =$ -0.25 THz using $\sigma^-$ polarisation. With these parameters we calculate the probability of infidelity during a $\pi$-pulse due to scattering from the intermediate state to be $1.4\times 10^{-4}$ using the open-source tool, Alkali Rydberg Calculator (ARC)~\cite{Sibalic:2017aa}. The total power efficiency between the lasers to atoms is currently limited by the use of an non-polarising beam splitter to combine the beams with the same polarisation before the fibre. This in future could be overcome by using lossless beam combining interferometers as presented in~\cite{Haubrich2000}.

\subsection{Rydberg Excitation Lasers}

As with the majority of neutral atom quantum processors published to date, interactions between atoms in our setup are enabled by exciting atoms to Rydberg states~\cite{Saffman:2010aa, Saffman_2016, Adams_2020, Morgado:2021aa, Graham:2022aa, Ebadi:2022aa}. This requires lasers locked to a high-finesse reference cavity to provide a stable frequency reference, whilst narrowing the laser sufficiently such that the laser linewidths are significantly smaller than the natural linewidths of the Rydberg states (typically of order of kHz)~\cite{Legaie:2018aa}.

Recent progress in neutral atom platforms has highlighted the need to suppress phase noise produced at higher frequencies (up to the MHz level) resulting from cavity locks in order to achieve high fidelity Rydberg operations~\cite{Levine:2018aa, Leseleuc:2018aa}. In this work, suppression of this high frequency noise is achieved using the low noise properties of the SolsTiS and locking to our commercially available high finesse cavity system developed for this purpose. This approach negates the requirement to inject a secondary laser using light filtered by the reference cavity~\cite{Levine:2018aa}. 

Atoms are excited to the 80$S_{1/2}$ Rydberg state using a two photon transition via 7$P_{1/2}$ with light at 1039~nm and 459~nm, Fig. 1(b). Over 2~W of light at 1039~nm is provided by a SolsTiS pumped by an Equinox and 2~W of light at 459 nm is generated by frequency doubling a 918~nm Equinox pumped SolsTis with our ECD-F module. The 1039~nm and 918~nm light are locked to our recently developed two-axis cavity system which has a free spectral range of 3 GHz and a cavity linewidth of ~ 75~kHz. The temperature of the cavity is stabilised via two feedback loops to an inner and outer radiation shielding layer and we maintain a temperature of $\pm$~4~mK. Using proprietary electronics, the lasers are locked with three feedback loops to an intracavity EOM, fast and slow piezos in a similar configuration to that used in our Raman phase locked system described earlier. 

Measurements of a beatnote with both lasers set to 918~nm and locked to independent cavity axes reveals a high suppression $>$ 40~dB of frequency noise beyond the laser linewidth, Fig. 1(c). Measuring the beatnote over 370 $\mu$s, we find an intrinsic laser linewidth (FWHM) of 10.4(2)~Hz, using a Gaussian fit.

\begin{figure}[t!]
\includegraphics{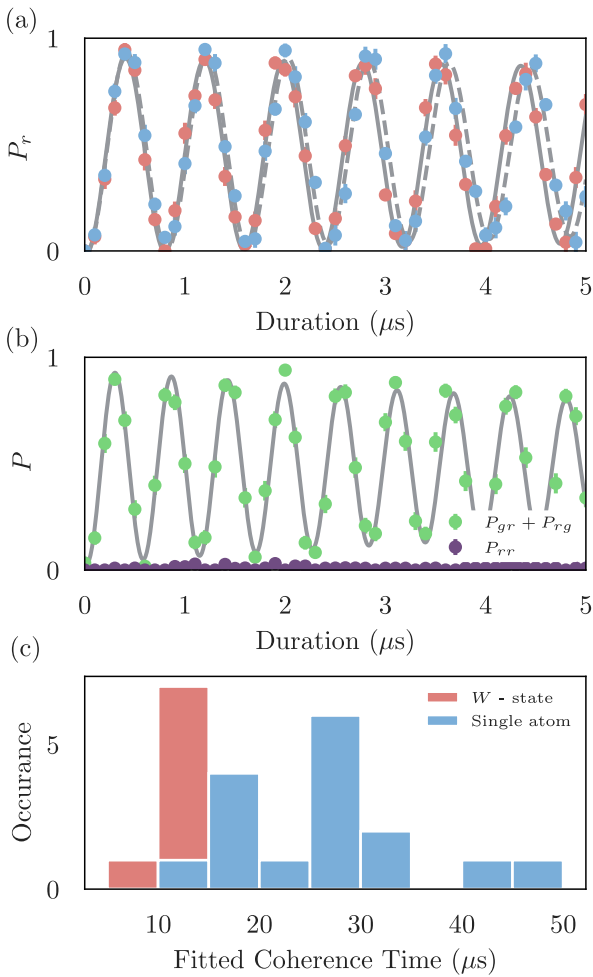}
\label{fig:fig4}
\caption{Rydberg Rabi oscillations of a single atom pair. (a) Single atom Rabi oscillations $\Omega_{a}/2\pi =$ 1.267(2) and 1.243(3)~MHz at $\Delta_{2}/2\pi =$~+~500MHz. (b) Rabi oscillation to the collective state when both sites load $\Omega_{W}/2\pi = 1.776(3)$~MHz~$\approx \sqrt{2}\Omega$, with high suppression to the double excited site $|rr\rangle$. (c) Histogram of the fitted coherence times to the Rabi oscillations across all sites/pairs, indicating high performance across the full array.}
\end{figure}

For the Rydberg experiments presented below, the 1039 nm laser is focused to a waist of 87~$\mu$m, with a power of 0.85 W delivered to the final focussing optics via fibre. The 459 nm light is focused to a $1/e^2$~waist of 105~$\mu$m and 65 mW is delivered to the experiment via fibre. Both lasers are intensity stabilised by feeding back a voltage from a photodiode after the fibre to an AOM. The laser intensity is sampled for a 1~ms period and the control value is held before extinguishing the light and opening shutters to the chamber for a Rydberg experiment to be conducted 2~ms later. This achieves an amplitude shot-to-shot stability of $\lesssim$~1~\% on both lasers. For all experiments presented, the single photon detuning of the Rydberg beams is $\Delta_{2}/2\pi =$  +500 MHz from the $7P_{1/2}, F=4$ state, with Rydberg Rabi frequencies of 1.25 MHz measured. Rydberg experiments are performed with the trapping potential turned off. The presence of a Rydberg atom is then verified through their loss as they are expelled from the trapping trap when the potential is turned back on.

\section{Coherent Control}

\subsection{Single Qubit Operations}

Coherent single qubit control is demonstrated on our 16 qubit sites. A variable duration Raman pulse is applied along a 7~G field to produce Rabi oscillations as shown in Fig.~3. We observe high contrast coherent oscillations on each site with coherence times beyond 50~$\mu$s and an average Rabi frequency of 1.58~MHz with a standard deviation across the sites of 0.5~\%. The ground state coherence time is measured via a Ramsey experiment and find $T_2^* =$8.8~ms averaged across the array with a standard deviation of 13~\%. This variation can be attributed to the differing trap depths across the sites limiting the final atomic temperature which has a direct correlation on $T_2^*$ and corresponds to an average atomic temperature of $\approx$~11.5~$\mu$K. 

Our optical pumping efficiency is measured by applying a slow Raman~$\pi$ pulse ($\Delta_1/2\pi =$ -1 THz, $\Omega/2\pi =$ 0.37 MHz) to minimise sensitivity to intensity noise from the Raman lasers. We measure a transfer probability between 94.1-98.5~\% with a mean across the array to be 97.1(7)~\%. Factoring in our previously discussed state measurement errors our optical pumping efficiency is found to be 98.1(7)~\%. This represents the largest source of state preparation and measurement (SPAM) infidelity in our setup. It is expected that utilising our Raman beams for Raman assisted optical pumping will increase our optical pumping efficiency to $>$~99~\% in the future~\cite{Levine:2019aa}. 

\subsection{High Fidelity Rydberg Control}

Coherent Rabi oscillations to $80S_{1/2}$ are demonstrated by varying the duration the Rydberg lasers are applied to our atoms pairs.
Due to the stochastic nature of how our pairs load we collect single atom and pair data demonstrating blockade and entanglement in the same experimental run. Our single atom data for two neighbouring sites is shown in Fig.~4(a) showing high contrast oscillations with minimal damping, indicating that detrimental phase noise from the cavity lock has been avoided. We observe a Rabi frequency of $\Omega/2\pi=$~1.267(2) and 1.243(3)~MHz and fitted 1/$e$ coherence times of 27(17) and 47(38)~$\mu$s. A slight asymmetry in the damping is observed which can be attributed to spontaneous emission from the intermediate state. From ARC~\cite{Sibalic:2017aa} we calculate the probability of this occurring to be 1.7\% per $\pi$ pulse.

Data from when each pair loaded is shown in Fig.~4(b), we observe the expected $\sqrt{2}$-enhancement in coupling to the excited symmetric $|W\rangle = (|gr\rangle+|rg\rangle)/\sqrt{2}$ state ($\Omega_{W}/2\pi = 1.776(3)$~MHz~$\approx \sqrt{2}\Omega$) and high suppression of the doubly excited state $|rr\rangle$, with the highest population measured in the first $2\pi$ of the rotation for the pair presented being 0.8(8)~\%. The coherence of this oscillation is shorter than our single atom oscillations which can be attributed to slightly different single atom Rabi frequencies between each atom in a pair and random Doppler shifts between each atom. To highlight our high fidelity control across the array we present a histogram of the fitted coherence times for the single atom and pair data in Fig. 4(c). We note that due to the oscillations only being measured out to 5~$\mu$s that there is a large error associated with the fit of the coherence and that this figure rather serves to demonstrate consistency of long coherence times across the array.

\begin{figure}[t!]
\includegraphics{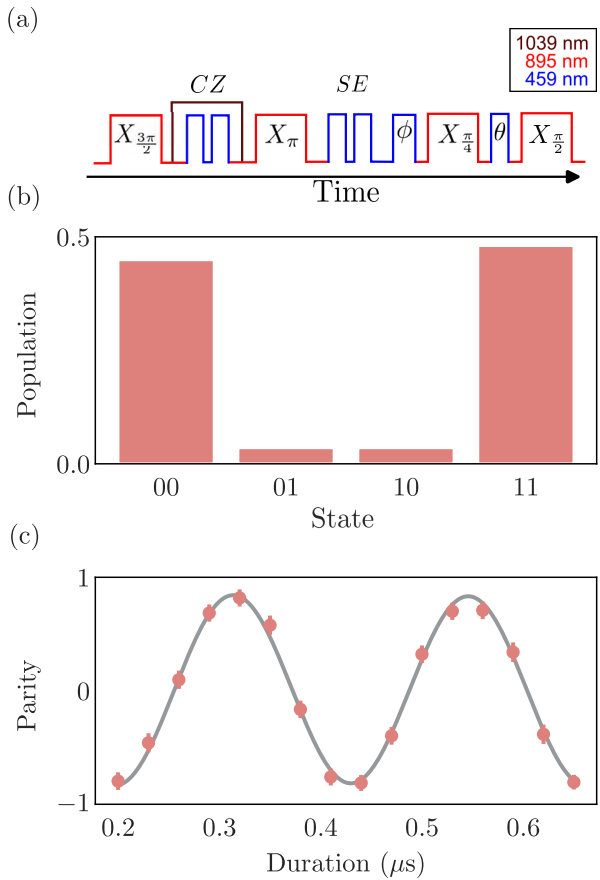}
\caption{(a) Pulse sequence used to create entangled Bell state $|\Psi^+\rangle $ and analysis pulses for parity oscillation. (b) Output populations for a single pair giving $P_{00}+P_{11} \ge 0.91(4)$. (c) Parity oscillation with contrast $C=0.83(2)$, giving a raw Bell state fidelity of $F\ge0.88(2)$.}
\label{fig:fig5}
\end{figure}

\subsection{Two-Qubit Gate and Entanglement}

We realise a two qubit gate using the protocol presented in~\cite{Levine:2019aa}, where two global Rydberg pulses of the equal duration are applied to create a CZ gate. Each pulse equates to a 2$\pi$ rotation via $|W\rangle$, thus the basis $|11\rangle$ returns to $|11\rangle$ with an acquired phase $\phi_{11}$ and $|00\rangle$ is unaffected. For basis $|01\rangle$ and $|10\rangle$ only one atom is excited to the Rydberg state, the first pulse makes an incomplete oscillation and the atom is then returned to the original basis state by varying the phase of the second pulse. The two pulses in the single atom case accumulate an additional phase $\phi_{01}$ and with the right choice in detuning, $2\phi_{01}-\pi = \phi_{11}$, realises a $CZ$ gate. In the regime where Rydberg blockade shift $V$ is much greater than the Rabi frequency, $\Omega \gg V$ this value is found to be $\Delta = 0.377\Omega$~\cite{Levine:2019aa, Graham:2022aa}. A final global phase shift using the 459~nm laser is then required to eliminate the single particle phase.

Local addressing to individually prepare our qubits in different basis states has not yet been implemented in our setup. However, the above gate protocol can be used to create an entangled state using only global pulses. A global $3\pi/2$ Raman pulse is applied and is followed by the gate protocol, with a final $\pi/4$ Raman pulse creating the Bell state $|\Psi^+\rangle =(|00\rangle+|11\rangle)/\sqrt{2}$. We also embed the CZ gate in an echo sequence as outlined in~\cite{Levine:2019aa}. Our pulse sequence can be seen in Fig. 5 (a). We measure the output populations, Fig. 5(b) and apply an analysis pulse using the 459~nm laser followed by a $\pi/2$ Raman pulse to induce a parity oscillation, shown in Fig. 5(c). From our best pair (Pair 1) we obtain populations $(P_{00}+P_{11}) = 0.93(4)$ and a parity oscillation amplitude contrast of $C=0.84(2)$, giving a raw Bell state fidelity of $F= (P_{00}+P_{11}+C)/2 =0.88(2)$. The average fidelity across our 7 best pairs is  $F=0.85$. 

Sources of error for the Bell state come from our previously discussed SPAM errors and atoms which are lost during the CZ protocol due to being left in the Rydberg state which leads to an overestimation of $P_{11}$. We repeat our measurement with no blowaway to determine the lower bound for our populations, following the analysis in~\cite{Levine:2019aa} we obtain $P_{00}+P_{11} \ge 0.91(4)$ giving a raw fidelity with a lower bound of $F\ge 0.875(20)$ (Pair 1). Finally we correct for our SPAM error, the probability of no error occurring for the pair of qubits is found to be $P=0.936(12)$, our corrected populations and parity oscillation amplitude are $P^c_{00}+P^c_{11} \ge 0.97(4)$  and $C^{c}=0.88(2)$ respectively giving a SPAM corrected fidelity with a lower bound of $F\ge 93(4)$. A summary of all 7 pairs is shown in Table 1. We note that we find minimal loss due to atoms being left in the Rydberg state, in Pair 1's case this error is 0.011(7), indicating excellent performance from our Rydberg lasers and that fidelities can be increased from addressing other experimental errors such as SPAM, homogeneity across the array and working at larger detuning from $7P_{1/2}$ to minimise the impact of spontaneous emission.

\begin{table}[t!]
\caption{\label{tab:table1}%
Analysis of Bell states on 7 pairs of atoms where $P$ is the probability of no error occurring for the pair of qubits, $P_{00}+P_{11}$ the lower bound populations, $C$ the parity oscillation contrast, $F$ the raw fidelity and  $F_c$ the SPAM corrected fidelity.}
\begin{ruledtabular}
\begin{tabular}{ lc c  c c c c}
\textrm{Pair}&
\textrm{$P$}&
\textrm{$P_{00}+P_{11}$}&
\textrm{$C$}&
\textrm{$F$}&
\textrm{$F_C$}
\\

\colrule
1& 0.936(12) &$\ge 0.92(4)$ & 0.83(2)&  $\ge 0.88(2)$ &$\ge 0.93(3)$ \\
2 & 0.946(12) &$\ge 0.91(4)$ & 0.82(3)&  $\ge 0.86(3)$ &$\ge 0.91(4)$ \\
3 & 0.937(12) &$\ge 0.87(4)$ & 0.74(5)&  $\ge 0.81(3)$ &$\ge 0.86(5)$ \\
4& 0.935(11) &$\ge 0.88(4)$ & 0.76(2)&  $\ge 0.82(2)$ &$\ge 0.87(4)$ \\
5 & 0.964(1) &$\ge 0.92(4)$ & 0.82(3)&  $\ge 0.87(3)$ &$\ge 0.90(3)$ \\
6 & 0.949(1) &$\ge 0.87(4)$ & 0.79(2)&  $\ge 0.83(2)$ &$\ge 0.87(3)$ \\
7 & 0.917(12) &$\ge 0.87(4)$ & 0.74(2)&  $\ge 0.81(2)$ &$\ge 0.87(3)$ \\

\end{tabular}
\end{ruledtabular}
\end{table}

\section{Conclusion}

We have demonstrated qubit operations using a photonics backbone of M Squared's lasers with extremely high atom-laser coherence times observed for Raman and Rydberg excitations. The lasers have also been used to create Bell states across 7 pairs of qubits with high fidelity. The leading causes for infidelities in our setup are due to the optical pumping efficiency, non-uniform trap depths, inhomogeneous addressing across the array and spontaneous emission from the intermediate state. Optical pumping efficiencies can be improved upon via Raman assisted optical pumping~\cite{Levine:2019aa}, the uniformity in trap depths at each qubit site can be improved on by feeding back to the SLM via techniques presented in~\cite{Schymik22} and uniform addressing can be improved upon by moving to flat top beams which has been demonstrated in~\cite{Ebadi:2021aa} allowing high fidelity operations in large scale 2D arrays. We note that the 459~nm ECD-F already provides sufficient power overhead (2~W) for moving to larger beams, with current internal investigation looking to increase available 1039~nm power which will also allow work at larger detuning reducing errors from spontaneous emission. Future work will look to introduce local operations with fast addressing such that two-qubit gates can be further benchmarked, and the introduction of atom sorting~\cite{Schymik22,Tian:2022aa} to allow investigation of larger 2D arrays and multi-qubit gates~\cite{Pelegri:2022aa}. We also note that the optimal power output of the SolsTiS, 780-820 nm, is ideal for future work in blue-detuned trapping of caesium~\cite{Graham:2019aa} or red-detuned trapping of rubidium~\cite{Ebadi:2021aa}.

\section{Acknowledgements}

The authors would like to thank the wider quantum team and staff at M Squared for their input and discussions. We would also like to thank the SQuAre team and Paul Griffin at the University of Strathclyde in their support for this work. This work is supported by Innovate UK, DISCOVERY (Project No: 50133). JP acknowledges support from EPSRC (Grant No. EP/T005386/1) and M Squared Lasers Ltd.

\end{document}